\documentclass[prd,aps]{revtex4}
\usepackage{graphicx}

\begin{document}
\title{Fundamental decoherence from relational time in 
discrete \\ quantum gravity: Galilean covariance}

\author{Rodolfo Gambini$^{1}$, Rafael A. Porto$^{2}$ and 
Jorge Pullin$^{3}$}
\affiliation {1. Instituto de F\'{\i}sica, Facultad de Ciencias,
Igu\'a 4225, esq. Mataojo, Montevideo, Uruguay. \\ 
2. Department of Physics, Carnegie Mellon University, 
Pittsburgh PA 15213\\
3. Department
of Physics and Astronomy, Louisiana State University,
Baton Rouge, LA 70803-4001}
\date{August 14th 2004}

\begin{abstract}
We have recently argued that if one introduces a relational time in
quantum mechanics and quantum gravity, the resulting quantum theory is
such that pure states evolve into mixed states.  The rate at which
states decohere depends on the energy of the states. There is
therefore the question of how this can be reconciled with Galilean
invariance. More generally, since the relational description is based
on objects that are not Dirac observables, the issue of covariance is
of importance in the formalism as a whole. In this note we work out an
explicit example of a totally constrained, generally covariant system
of non-relativistic particles that shows that the formula for the
relational conditional probability is a Galilean scalar and therefore
the decoherence rate is invariant.
\end{abstract}

\maketitle

\section{Introduction}

We have recently introduced a new technique for discretizing physical
theories \cite{DiGaPu}. When applied to general relativity it yields a
discrete theory that is constraint-free yet it approximates well the
continuum theory under certain circumstances \cite{GaPu,cosmo}.  The
lack of constraints allows to tackle some of the fundamental open
problems of canonical quantum gravity. For instance one can introduce
a relational time \cite{greece,deco1,njp} a la
Page--Wootters \cite{PaWo}. That is, one promotes all quantities in
the theory to quantum operators and chooses one that is called
``clock'' and then one computes conditional probabilities for the other
variables to take given values when the ``clock'' variable shows a
certain ``time''. The resulting quantum theory approximates ordinary
quantum mechanics well when the clock variable chosen behaves in a
semi-classical fashion with small quantum fluctuations.  If one
chooses as clock a variable that is in a quantum regime, the resulting
theory is still valid but it will not resemble ordinary quantum
mechanics.

We have also argued that, due to the fact that one cannot have a 
perfectly classical clock in nature, the resulting theory will have
small but non-vanishing departures from ordinary quantum mechanics.
In particular a pure state does not remain pure forever but evolves
into a mixed state.

Since one is approximating a constrained continuum theory with a
discrete theory that is unconstrained, the resulting relational
discrete theory is formulated in terms of variables that are
observables for the discrete theory. But they are not necessarily the
discrete counterpart of Dirac observables of the continuum theory.
Therefore the issue of how to reconcile the predictions of the
discrete relational theory with the covariance of the continuum theory
is of importance. In particular, the conditional probabilities must
remain invariant when one changes coordinates and both the clock
variable and the observed variable change values. To tackle the
covariance problem in complete generality is beyond the scope of this
paper.  What we intend to do here is to analyze a simple model where
calculations can be worked out concretely and in particular to probe
the following issue. Since the prediction for the time of decoherence
of pure states results in a formula that involves the energy of the
states, it is may not be immediately apparent in what sense it is
Galilean invariant.  We would like to discuss in a simple model how to
interpret the formula in a way that the invariance is manifest.

The organization of this paper is as follows. In section II we present
the model we will study and in section III we will show the emergence
of Galilean invariance. We end with a discussion.

\section{The model}

We consider the following model. It consists of two non-interacting
particles moving in separate potentials in $1+1$ dimensions. One of
the particles we will assume is much more massive than the other and
it will determine the variable we choose as a clock. The other
particle we will call the ``system'' particle. The potential affecting
the clock particle will be a constant force field. We will assume the
particle is far away from the turning point, since in this regime we
know our discrete approach approximates well the continuum
\cite{cosmo}. For the system particle we will assume it behaves
quantum mechanically and is in a potential that gives rise to bound
states. As is well known \cite{brho}, the best way to understand
Galilean transformations in quantum mechanics is to study them as a
limit of Lorentz transformations. We will therefore choose Lorentz
invariant and reparameterization invariant action for the particles,
\begin{equation}
S =\int d\tau \left[
-\left(M c +{U(q^0,q)\over c}\right)\sqrt{(\dot{q}^0)^2-\dot{q}^2}
-\left(m c+ {V(\phi^0,\phi)\over c}\right)\sqrt{(\dot{\phi}^0)^2-\dot{\phi}^2}\right]
\end{equation}
where $q^0,q$ are the space-time coordinates of the ``clock'' particle
and $\phi^0,\phi$ are the space-time coordinates of the ``system''
particle. We have kept the speed of light explicit in order to
consider later the non-relativistic limit. We start by assuming that
the time-like coordinates of both particles can be synchronized (since
we will work in the Newtonian limit this poses no conceptual problem)
$q^0=\phi^0$, and the reference system has been chosen such that
$\dot{q}^0\gg \dot{q}$ and $\dot{\phi}^0\gg \dot{\phi}$, and also
$Mc^2\gg U(q^0,q)$ and $mc^2\gg V(\phi^0,\phi)$ (non-relativistic limit).  
For concreteness we
assume that in the reference frame given, $U(q^0,q)=\alpha q$ and
$V(\phi^0,\phi)=V(\phi)$ and the latter has bound states.  With these
assumptions the action becomes,
\begin{equation}
S =\int d\tau \left[
-M c \dot{q}^0 -{\alpha q \over c} \dot{q}^0+M c {\dot{q}^2 \over 2
\dot{q}^0} -m c \dot{\phi}^0-{V(\phi)\over c} \dot{\phi}^0+
Mc {\dot{\phi}^2 \over 2 \dot{\phi}^0}+\lambda(q^0-\phi^0)\right]
\end{equation}
where $\lambda$ is a Lagrange multiplier associated with the
constraint that imposes the synchronization. It is immediate to see
that if one chooses $q^0=\phi^0=ct$ one will obtain the ordinary
action for two non-relativistic particles with $t$ the ordinary
non-relativistic time. We will not do this here, since we are
interested in handling a totally constrained system, since it is in
such systems where the introduction of a relational time is meaningful
since they have no preferred notion of time.

To understand better the constraint structure of the theory, we 
will rewrite the action in first-order form. We define the 
canonical momenta,
\begin{eqnarray}
p_0 &=& {\partial L \over \partial \dot{q}^0} =-M c-{\alpha q \over c} 
-{M c \over 2} {\dot{q}^2 \over (\dot{q}^0)^2},\\
p &=& {\partial L \over \partial \dot{q}} = M c{\dot{q}\over
\dot{q}^0},\\
\pi_0&=&{\partial L \over \partial \dot{\phi}^0} =-m c-
{V(\phi) \over c} -{M c \over 2} {\dot{\phi}^2 \over (\dot{\phi}^0)^2},\\
\pi &=& {\partial L \over \partial \dot{\phi}} = m {\dot{\phi}\over
\dot{\phi}^0}.
\end{eqnarray}

From where we can get two constraints, in addition to the one
we had before $\phi^0-q^0=0$,
\begin{eqnarray}
p_0&= & -M c-{\alpha q\over c} -{p^2 \over 2 Mc}= -{1 \over c}
H_1(p,q),\\ \pi_0 &=&-m c -{V(\phi)\over c} - {\pi^2 \over 2 mc}=-{1
\over c} H_2(\phi,\pi).
\end{eqnarray}

If we rearrange the latter two constraints into their sum and
difference,
\begin{eqnarray}
\pi_0+p_0 &+& {H_1 \over c} +{H_2 \over c}=0, \label{conssum}\\
\pi_0-p_0 &+& {H_1 \over c} -{H_2 \over c}=0, 
\end{eqnarray}
one readily sees that the last constraint together with 
$q^0-\phi^0=0$ are second class, whereas they both commute with
(\ref{conssum}). One imposes the second class constraints strongly
and is left with a theory with one constraint, whose action is,
\begin{eqnarray}
S&=&\int d\tau \left(\left(p_0+\pi_0\right)\dot{q}^0
+p \dot{q} +\pi \dot{\phi} + N \left[p_0+\pi_0
+{H_1 \over c} +{H_2 \over c}\right]\right)\\
&=&\int d\tau \left(\tilde{p}_0\dot{q}^0
+p \dot{q} +\pi \dot{\phi} + N \left[\tilde{p}_0
+{H_1 \over c} +{H_2 \over c}\right]\right)
\end{eqnarray}
where we introduced the shorthand $\tilde{p}_0\equiv p_0+\pi_0$. This
action is very natural for the system under study (in fact we could
have started the calculation simply by considering this action from
the outset).

\section{Galilean invariance}

To probe the invariance of the decoherence effect of interest, we
would like to study two different situations. One in which the system
particle is a potential $V(\phi)$ and another in which the potential
is of the form $V(\phi-\beta q^0)$, this will represent a system that
is bound by a potential around some minimum that is fixed or that is
moving, respectively. This corresponds to adopting the active point of
view of the Galilean transformation. We will do this with our
consistent discretization techniques \cite{GaPu,DiGaPu,cosmo}. We
refer the reader to our previous papers for details on the technique.
We start by discretizing the action in the first of the two cases of
interest. The integral in the action becomes replaced by a discrete
sum $S=\sum_0^N L(n,n+1)$ and we absorb the time interval
$\epsilon=\tau_{n+1}-\tau_n$ and where,
\begin{equation}
L(n,n+1) = \tilde{p}_{0n} \left(q^0_{n+1}-q^0_n\right)
+p_n \left(q_{n+1}-q_n\right) +\pi_n \left(\phi_{n+1}-\phi_n\right)
 - N_n \left[\tilde{p}_{0n}
+{H_1(p_n,q_n) \over c} +{H_2(\pi_n,\phi_n) \over c}\right].
\end{equation}
We now implement the canonical transformation that materializes
the time evolution between instant $n$ and $n+1$ with the
Lagrangian $-L(n,n+1)$ playing the role of generating function of
a type I canonical transformation,
\begin{eqnarray}
P^{\tilde{p}_0}_{n+1} &=& 
{\partial L(n,n+1) \over \partial \tilde{p}_{0n+1}} =0,\\
P^{\tilde{p}_0}_n &=& -{\partial L(n,n+1) \over 
\partial \tilde{p}_{0n}} = -\left(q^0_{n+1}-q^0_n\right) +N_n,\\
P^{q^0}_{n+1} &=& 
{\partial L(n,n+1) \over \partial {q}^0_{n+1}} 
=\tilde{p}_{0n},\\
P^{q^0}_{n} &=& 
-{\partial L(n,n+1) \over \partial {q}^0_{n}} 
=\tilde{p}_{0n},\\
P^{p}_{n+1} &=& 
{\partial L(n,n+1) \over \partial p_{n+1}} =0,\\
P^{p}_{n} &=& 
-{\partial L(n,n+1) \over \partial p_{n}} =
-\left(q_{n+1}-q_n\right) +N_n{p_n \over  M},\\
P^{q}_{n+1} &=& 
{\partial L(n,n+1) \over \partial q_{n+1}} =p_n,\\
P^{q}_{n} &=& 
-{\partial L(n,n+1) \over \partial q_{n}} =
p_n+\alpha N_n,\\
P^{\pi}_{n+1} &=& 
{\partial L(n,n+1) \over \partial \pi_{n+1}} =0,\\
P^{\pi}_{n} &=& 
-{\partial L(n,n+1) \over \partial \pi_{n}} =
-\left(\phi_{n+1}-\phi_n\right)-N_n{\pi_n \over m},\\
P^{\phi}_{n+1} &=& 
{\partial L(n,n+1) \over \partial \phi_{n+1}} =\pi_n,\\
P^{\phi}_{n} &=& 
-{\partial L(n,n+1) \over \partial \phi_{n}} =
\pi_n+N_n{\partial V(\phi_n) \over \partial \phi_n},\\
P^{N}_{n+1} &=& 
{\partial L(n,n+1) \over \partial N_{n+1}} =0,\\
P^{N}_{n} &=& 
-{\partial L(n,n+1) \over \partial N_{n}} =
\tilde{p}_{0n}
+Mc +\alpha {q_n\over c} + {(p_n)^2 \over 2 M c}
+mc +{V(\phi_n)\over c} + {(\pi_n)^2 \over 2 m c}.
\end{eqnarray}

The system has constraints and we will use them to eliminate
some of the variables and yield a system of evolution equations
in a more explicit form. The resulting system is the following,
\begin{eqnarray}
q^0_{n+1}&=&q^0_n+N_n,\\
P^{q^0}_{n+1} &=& P^{q^0}_n,\\
q_{n+1}&=&q_n+N_n {P^q_{n+1} \over M},\\
P^q_{n+1}&=& P^q_n-\alpha N_n,\\
\phi_{n+1} &=& \phi_n+N_n {P^\phi_{n+1}\over m},\\
P^\phi_{n+1} &=& P^\phi_n-N_n {\partial V(\phi_n)\over \partial
\phi_n},\\
0&=&P^{q^0}_{n+1}
+Mc +\alpha {q_n\over c} + {(P^q_{n+1})^2 \over 2 M c}
+mc +{V(\phi_n)\over c} + {(P^\phi_{n+1})^2 \over 2 m c}.
\label{34}
\end{eqnarray}

The last equation  determines
the Lagrange multiplier $N_n$. To see this, we first rewrite it
entirely in terms of variables at $n+1$,
\begin{equation}
P^{q^0}_{n+1}
+Mc +\alpha {{q_{n+1} -N_n P^q_{n+1}}\over Mc} + 
{(P^q_{n+1})^2 \over 2 M c}
+mc +{V({\phi_{n+1}-N_n P^\phi_{n+1}})\over mc} + 
{(P^\phi_{n+1})^2 \over 2 m c}=0.
\end{equation}

Since we are ultimately interested in studying the system
in a regime close to the continuum limit, we make the assumption
that the lapse $N_n$ is small and expand the term involving the
potential to first order in $N_n$, 
\begin{equation}
P^{q^0}_{n+1}
+Mc +\alpha {{q_{n+1} -N_n P^q_{n+1}}\over Mc} + 
{(P^q_{n+1})^2 \over 2 M c}
+mc +{V(\phi_{n+1})\over mc}  
-{N_nP^\phi_{n+1}\over mc}
{\partial V(\phi_{n+1})\over \partial \phi_{n+1}}
+{(P^\phi_{n+1})^2 \over 2 m c}=0.
\end{equation}

We can now solve explicitly for the Lagrange multiplier,
\begin{equation}
N_n=\left(\alpha {P^q_{n+1} \over Mc} +{P^\phi_{n+1} \over mc} V'(\phi_n)\right)^{-1}C_{n+1}
\end{equation}

Where $C_{n+1}$ is the constraint of the continuum theory 
discretized as if all variables were at $n+1$,
\begin{equation}
C_{n+1} = P^{q^0}_{n+1}
+Mc +\alpha {q_{n+1}\over c} + {(P^q_{n+1})^2 \over 2 M c}
+mc +{V(\phi_{n+1})\over c} + {(P^\phi_{n+1})^2 \over 2 m c}.
\end{equation}

We now assume that $\alpha \gg V'(\phi)$. This is due to the fact
that we are assuming the clock to be classical and large and $\alpha$
is therefore associated with a macroscopic force whereas $V(\phi)$ is
the potential in which the system is bound, and the latter is 
microscopic in nature. With this assumption we make sure there
are no singularities in the computation of the Lagrange multiplier.
Recall that the discrete description departs from the continuum one
close to the turning point of the orbit.

One can now substitute the expression of the Lagrange multiplier in
the evolution equations. The resulting system of equations can be
viewed as a canonical transformation between instant $n$ and instant
$n+1$ for the remaining variables of the problem.  The next step
consists in quantizing the system by representing the discrete
evolution through a unitary operator, i.e.
$\hat{z}^i_n=\hat{U}^\dagger \hat{z}^i_{n+1} \hat{U}$ where the
$z^i$'s are all the phase space variables of the problem. All these
calculations can be worked out explicitly for a simple system like the
one we are considering, we will not show all the details here for
reasons of space, the reader can see similar treatments in
\cite{greece,smolin}.

Since we are interested in the continuum limit, a shortcut can be
taken by considering the Hamiltonian associated with the unitary
transformation $\hat{U}=e^{i\hat{H}}$ \cite{smolin}.  The Hamiltonian
is obtained by taking the logarithm of the unitary operator as a power
series. This power series is convergent at all points in phase space
except for a small region around the turning point of the orbit of the
clock system. The Hamiltonian is obviously conserved upon evolution
(except at the turning point). The first term in the expansion of the
Hamiltonian is,
\begin{equation}
H_n \sim {M c (C_n)^2 \over \alpha p_n}\left[1 +O\left(
{M c\, C_n\over (p_n)^2}\right)\right],
\end{equation}
and to simplify notation, from now on we call $P^q_n=p_n$ and 
$P^{q^0}_n=p_{0n}$.

For the quantization we consider wavefunctions $\psi_n(q^0,q,\phi)$
forming a Hilbert space at the ``instant'' $n$. Isomorphic
Hilbert spaces exist at all other discrete instants.
With the Hamiltonian we will study the evolution operator
$\hat{U}(n,n_0) =e^{i\hat{H} (n-n_0)}$ and its action on the states,
$\psi_n(q^0,q,\phi)=\hat{U}(n,n_0)\psi_{n_0}(q^0,q,\phi)$, and we are
working in the Schr\"odinger representation. The explicit form of 
the quantum Hamiltonian is,
\begin{equation}
\hat{H}={M c \over \alpha \hat{p}}\left(\hat{p}_0
+\hat{H}_1+\hat{H_2}\right)^2.
\end{equation}

It is to be noted that the expression in parenthesis is the 
constraint that one has in the continuum theory. In the consistent
discretization approach the constraint of the continuum theory is not
enforced exactly (what is enforced is equation (\ref{34}) which 
corresponds to the constraint of the continuum theory but with the
momenta evaluated one instant after the configuration variables).
In the continuum limit, it nevertheless is enforced quite approximately
and therefore the norm of $\hat{H}$ is going to be small.

We consider a quantum state in which the clock has a semiclassical
behavior, so it is described by a coherent state peaked at
$<\hat{H}_1>_{n_0}=\bar{E}$, $<\hat{q}_0>_{n_0}=0$,
$<\hat{q}>_{n_0}=\bar{q}$, $<\hat{p}>_{n_0}=\bar{p}$ and $
<\hat{p}_0>_{n_0}=\bar{p}_0$. We then have for the wavefunction,
\begin{equation}
\Psi_{n_0} = \psi_{n_0}(q,q^0) \varphi_{n_0}(\phi) 
\end{equation}
with
\begin{equation}
\psi_{n_0}(q,q^0)=\left(2\pi\sigma_1^2\right)^{-1/4}
\exp\left[-\frac{\left(q-\bar{q}\right)^2}{4 \sigma_1^2}+i \bar{p} q\right]
\left(2\pi\sigma_0^2\right)^{-1/4}
\exp\left[-\frac{\left(q^0\right)^2}{4 \sigma_0^2}+i \bar{p}_0 q^0\right]
\end{equation}
where $\sigma_1$ is the dispersion in the variable $q$ and $\sigma_0$
is the dispersion in the variable $q^0$. We have also assumed that
$\bar{E}\gg |\bar{p}_0+\bar{E}| \gg\, <\hat{H}_2>_{n_0}$. The first
inequality is in order to be in the continuum limit. The second
inequality is in order to simplify calculations, and implies that we
are accepting as ``continuum limit'' a regime where the constraint of
the continuum theory is well enforced with respect to the scale of
energies of relevance for the ``clock'' system, but the error in 
enforcement is large with respect to the energies of the system under
study. It would be desirable to extend the results of this paper to 
regimes that approximate even further the continuum theory, but the
calculations would be more involved.

The fundamental equation to be studied is the conditional
probability,
\begin{equation}
P(\phi \in \Delta \phi|q^0 \in \Delta t) = 
{\sum_n {\rm Tr}\left(\hat{U}^\dagger(n) \hat{P}_{\phi,{q}^0} \hat{U}(n)
\rho_{q^0}\times\rho_q \times \rho_\phi\right)
\over \sum_n {\rm Tr}\left(\hat{U}^\dagger(n) \hat{P}_{q^0} \hat{U}(n)
\rho_{q^0}\times\rho_q \times \rho_\phi\right)}
\end{equation}
with $\hat{P}_{\phi,q^0}$ is the projector onto the eigenstate
labeled by the values $\phi,q^0$ and $\rho_{q^0}, \rho_q,\rho_\phi$
the density matrices associated with the state $\Psi_{n_0}$. 
From now on we will use natural units where $c=\hbar=1$.

Let us analyze the denominator of this expression. Taking the 
trace on the $\phi,q$ spaces by integrating, we get,
\begin{eqnarray}
{\rm Den}&=&\sum_n {\rm Tr}\left(\hat{U}^\dagger(n) \hat{P}_{q^0} \hat{U}(n)
\rho_{q^0}\times\rho_q \times \rho_\phi\right)\\ &=& 
\sum_n {\rm Tr}\left[\exp\left(-i\frac{\left(\hat{p}_0
+\bar{E}\right)^2}{\frac{\alpha \bar{p}}{M}}\left(
n-n_0\right)-2i\frac{\hat{p}_0 +\bar{E}}{\frac{\alpha
\bar{p}}{M}}<\hat{H}_2>\left(n-n_0\right)\right)\times\right.\nonumber\\
&\times&\left.\hat{P}_{q^0} 
\exp\left(i\frac{\left(\hat{p}_0
+\bar{E}\right)^2}{\frac{\alpha \bar{p}}{M}}\left(
n-n_0\right)+2i\frac{\hat{p}_0 +\bar{E}}{\frac{\alpha
\bar{p}}{M}}<\hat{H}_2>\left(n-n_0\right)\right)
\rho_{q^0}\right],\nonumber
\end{eqnarray}
and the term involving $\hat{H}_2^2$ from $U$ cancels with that of
$\hat{U}^\dagger$ since $\hat{H}_2^2$ commutes with $\hat{P}_{q^0}$.
We have also replaced $\hat{H}_1$ by $\bar{E}$ and $\hat{p}$ by
$\bar{p}$ since the trace implies taking the expectation value of
quantities depending on $q$ and $p$.  Since $\rho_{q^0}$ represents
a state very peaked at $<\hat{p}_0>=\bar{p}_0$ and $<\hat{q}^0>=0$ and
since $|\bar{p}_0+\bar{E}|\gg\,<\hat{H}_2>$, we have that,
\begin{equation}
{\rm Den} = \sum_n {\rm Tr}\left[
\hat{P}_{q^0} 
\exp\left( i \frac{\left(\hat{p}_0+\bar{E}\right)^2}
{\frac{\alpha \bar{p}}{M}}(n-n_0)^2\right) 
\rho_{q^0}
\exp\left( -i \frac{\left(\hat{p}_0+\bar{E}\right)^2}
{\frac{\alpha \bar{p}}{M}}(n-n_0)^2\right) \right] 
\equiv 
\sum_n {\rm Tr}\left(\rho_n(q^0)\right)\label{45}
\end{equation}
where $\rho_n(q^0)\equiv \hat{P}_{q^0} \rho_{n,q^0}\equiv
  \hat{P}_{q^0} \hat{U}(n)\rho_{q^0} \hat{U}^\dagger(n)$ represents
  the wavepacket of a ``free particle'' which evolves with the
  effective Hamiltonian
\begin{equation}
\hat{H}_{\rm eff} ={(\hat{p}_0+\bar{E})^2\over 
{\alpha \bar{p} \over M}}
\end{equation}

It is instructive to realize that one can write,
\begin{equation}
{\rm Tr}\left[\rho_n(q^0)\right] = 
\left(2 \pi \sigma_0^2(n)\right)^{-1/4} \exp \left[-\frac{(q^0
-\bar{q}_0(n))^2}{4 \sigma_0^2(n)}\right],
\end{equation}
where
\begin{equation}
\bar{q}^0(n) = 2 {(\bar{p}_0+\bar{E})(n-n_0)\over {\alpha \bar{p}\over M}}
\equiv t_{\rm max}(n),\label{48}
\end{equation}
which shows that the clock ``displays a time'' in the neighborhood
of $\bar{q}^0$ when we are 
at the level $n$ of the discrete theory. We have defined $t_{\rm
max}(n)$,
the most likely value of the clock ``time'' for a given $n$ level
in the discrete theory, and we have chosen the clock in such a
way that $t_{\rm max}$ grows linearly with $n$.

The width of the packet grows with $n$ as,
\begin{equation}
\sigma_0^2(n)=\sigma_0^2\left(1 + \frac{1}{4 \sigma_0^4} \left({M \over
\alpha \bar{p}}\right)^2(n-n_0)^2\right).\label{49}
\end{equation}

We now should introduce some relevant scales. We will assume the
characteristic mass of the clock system is about a kilogram. The
potential of the clock system is characterized by the macroscopic
constant $\alpha$, which we will assume is of the order of 10 Newton,
which in natural units corresponds to $\alpha \sim 10^{22} m^{-2}$,
which implies that if we have $\sigma_0 \sim 10^{-10} s \sim 1 m$,
then,
\begin{equation}
\frac{1}{4 \sigma_0^2}\left({M \over \alpha \bar{p}}\right)^2 \sim
10^{-34} m^2,
\end{equation}
and we have assumed $\bar{p}/M\sim 10^{-5}$ so we are in a non-relativistic
regime.

As we discussed in \cite{greece}, the sums that appear in the
numerator and denominator for the conditional probability should be
large enough to involve the complete evolution of interest for the
system, but they should not be infinite, otherwise one gets an
indeterminate quotient of two diverging quantities for the conditional
probability. Given the value computed above for the quantity
multiplying $(n-n_0)^2$, it is natural to bound the value of $n-n_0\ll
10^{17}$, that is we assume that the sums go from $n_0$ to a maximum
value $N\ll 10^{17}$, let's say $N\sim 10^{14}$, otherwise the packet
representing the clock will spread too much and we would be out of the
semiclassical regime. Notice that we also have that $\bar{E}\geq
10^{26} m^{-1}$, and recalling that $|\bar{p}_0+\bar{E}|$ has to be
smaller than $\bar{E}$, and choosing it to be $10^{17} m^{-1}$ yields
$\bar{q}^0\sim 10^4 s\sim 3$ hours, which is a reasonable number.
Summarizing, by bounding the number of steps we find that the
denominator is a quantity of order unity. Its precise value is not of
great interest, since we can choose it by fixing the normalization of
the probability.

Let us analyze the numerator,
\begin{equation}
\label{numerator}
{\rm Numer} = \sum_n {\rm Tr} \left[ \hat{P}_{\phi,q^0}
\exp\left( i {(\hat{p}_0+\bar{E})^2 +2 (\hat{p}_0+\bar{E})\hat{H}_2
+\hat{H}_2^2\over {\alpha \bar{p}\over M}}\right)
\rho_{q^0} \rho_q \rho_\phi 
\exp\left( -i {(\hat{p}_0+\bar{E})^2 +2 (\hat{p}_0+\bar{E})\hat{H}_2
+\hat{H}_2^2\over {\alpha \bar{p}\over M}}\right)\right],
\end{equation}
and observing that the projector is independent of $q$, and 
one can therefore substitute $H_1$ by its expectation value
$\bar{E}$. Using now that the clock is semiclassical and
$\bar{p}_0+\bar{E} \gg\, <\hat{H}_2>$ to neglect terms quadratic in
$\hat{H}_2$ we can write,
\begin{eqnarray}
P(\phi \in \Delta \phi| q^0 \in \Delta t) &=&
\sum_n {\rm Tr}\left[
\hat{P}_{\phi,q^0} 
\exp\left(i {\left[(\hat{p}_0+\bar{E})^2 +2
(\bar{p}_0+\bar{E})\hat{H}_2\right](n-n_0) \over {\alpha \bar{p}\over
M}}\right) 
\rho_\phi(n_0)\rho_{q^0}(n_0)\right.\nonumber\\
&\times&\left.\exp\left(-i {\left[(\hat{p}_0+\bar{E})^2 +2
(\bar{p}_0+\bar{E})\hat{H}_2\right](n-n_0) \over {\alpha \bar{p}\over
M}}\right) \right] {\rm Den}^{-1}\\
&=& \sum_n {\rm Tr}\left[
\hat{P}_\phi 
\exp\left(i\hat{H}_2 t_{\rm max}(n)\right)
\rho_\phi \exp\left(-i\hat{H}_2 t_{\rm max}(n)\right)\right] \times
\nonumber\\
&\times& 
{\rm Tr} \left[\hat{P}_{q^0} 
\exp\left(i {(\hat{p}+\bar{E})^2 \over 
{\alpha \bar{p} \over M}}(n-n_0)\right)
 \rho_{q^0}
\exp\left(-i {(\hat{p}+\bar{E})^2 \over
{\alpha \bar{p} \over M}}(n-n_0)\right)\right] {\rm Den}^{-1},
\end{eqnarray}
where we have replaced $\hat{p}_0$ with $\bar{p}_0$ since the clock is
approximately classical and its energy dominates in $\bar{p}_0$, and
we have separated the expression into two pieces, one dependent on the
$\phi$ variable and one dependent on the $q_0$ variable to make more
explicit the separation between clock and system.

The last trace divided by ${\rm Den}$ 
can be written as ${\cal P}_n(q^0)$ and satisfies that
$\sum_n {\cal P}_n(q^0)=1$.

Following the discussion in \cite{njp}, in order to make contact
with ordinary quantum mechanics we assume the spacing in $n$ is 
small compared with the values of $n$ and introduce a continuous
variable $v=n \epsilon$. We choose $\epsilon$ such that 
\begin{equation}
\epsilon = 2 {\bar{p}+\bar{E} \over {\alpha \bar{p} \over \bar{E}}},
\end{equation}
so we have that $\epsilon\le 1 m$ with the choice of scales we made
for the problem. We choose $n_0=0$ and we can then write a good
continuum limit approximation for ${\cal P}_n(q^0)$, as in \cite{njp},
\begin{equation}
{\cal P}_v(q^0) = \delta(v-q^0)+\sigma_0^2(q^0) \delta''(v-q^0),
\end{equation}
with $\sigma_0^2(q^0)$ given by (\ref{48},\ref{49}),
\begin{equation}
\sigma_0^2(q^0) = \sigma_0^2 \left(1 + \frac{1}{4 \sigma_0^4} 
\frac{(q^0)^2}{4 (\bar{p}_0+\bar{E})^2}\right)
\end{equation}
and with $t_{\rm max}(n)=\epsilon n =v$, so we can write,
\begin{equation}
P(\phi \in \Delta \phi|t \in \Delta t) = \int dv {\rm Tr}\left[\hat{P}_\phi
\exp\left(i \hat{H}_2 v\right) 
\rho_\phi 
\exp\left(-i \hat{H}_2 v\right) \right] \left(\delta(v-q^0)+
\sigma_0^2(q^0) \delta''(v-q^0)\right)={\rm
Tr}\left[\tilde{\rho}_2(q^0) \hat{P}_\phi\right]
\end{equation}
where
\begin{equation}
\tilde{\rho}_2(q^0)= \int dv  {\cal P}_v(q^0) 
\hat{U}_v \rho_\phi(v=0) \hat{U}^\dagger_v,
\end{equation}
and this density matrix satisfies a Schr\"odinger equation modified due
to the fact that we are considering a quantum clock as shown in detail in \cite{njp},
\begin{equation}
{\partial \tilde{\rho}_2 \over \partial q^0} = 
i[\hat{H}_2,\tilde{\rho}_2] - \sigma(q^0) 
[\hat{H}_2,[\hat{H}_2,\tilde{\rho}_2]], \label{59}
\end{equation}
with $\sigma(q^0) = d\sigma_0^2(q^0)/d q^0$. This expression is just
the first two terms in a power series in terms of the dispersion of the
quantum clock, which for realistic systems is a very small quantity.

To try to get a handle on a rough value for this quantity in the case
of a realistic system, we note that the macroscopic clock particle is
subject to decoherence due to interaction with the environment. If we
characterize such decoherence by a time $t_D$, we have,
\begin{equation}
\sigma \sim \frac{1}{4 \sigma_0^2} \frac{q^0}
{2 (\bar{p}_0+\bar{E})^2}|_{q^0=t_D}.
\end{equation}

If $t_D\sim 1s \sim 10^{10}m$, which is a rather large decoherence
time for a macroscopic system, then $\sigma \sim 10^{-24}m$. In 
reference \cite{bh} we have estimated theoretical limits as to how
small a dispersion is attainable with optimal realistic clocks.

In order to study the Galilean covariance of the conditional
probability, the procedure is simple.  We have to repeat the
calculation assuming a boost with velocity $-\beta$ has been performed
on the system 2 respect to the system 1, in such a way that the
potential it now sees is of the form $V(\phi-\beta q^0)$. For
instance, the system 2 can be an electron in a central potential given
by a nucleus. The relational analysis goes along exactly as before,
with two differences. The Hamiltonian for the second system becomes,
\begin{equation}
H'_2 = V(\phi-\beta q^0)+{\pi^2 \over 2 m}, 
\end{equation}
and the initial state of the system is given by 
$\rho_{q^0}'\times \rho_\phi' =\hat{U}_G\left(\rho_{q^0}\times \rho_\phi\right) \hat{U}_G^\dagger$ with
\begin{equation}
  \label{62}
  \hat{U}_G = \exp\left[i \beta \hat{\pi} \hat{q}^0 -i m \beta \hat{\phi}\right].
\end{equation}
In other words, the initial state is the one corresponding to the
Galilean boost $\hat{U}_G$ to the original state \cite{brho}. Notice
however that in traditional treatments of Galilean invariance in 
quantum mechanics the variable that we here take as $\hat{q}^0$ is 
a classical parameter $t$. Our treatment can be considered a relational
generalization of the usual Galilean transformations of quantum mechanics.
In ordinary quantum mechanics Schr\"odinger's equation has a time derivative
that acts on the parameter in $\hat{U}_G$. In the relational treatment
the equation has a term involving $\hat{p}_0$ instead of the time derivative.
Notice that $\hat{p}_0$ is minus the total energy instead of just the
``system energy''. 
Therefore the presence of the operator $\hat{q}^0$ in $\hat{U}_G$ has 
the same effect in the relational treatment as the derivative with 
respect to the parameter has in ordinary quantum mechanics: they both
induce a change in the energy of the system due to the boost, $\hat{p}_0\rightarrow \hat{p}_0+\beta \hat{\pi}$.

To study the changes in the conditional probability we go back in the
derivation to equation (\ref{numerator}),
\begin{eqnarray}
  P'\left(\phi \in \Delta \phi | q^0 \in \Delta t\right) &=& 
\sum_n {\rm Tr}\left[ 
\hat{P}_{\phi,q^0} \exp\left(i{\left[\left(\hat{p}_0+\bar{E}\right)^2+
2 \left(\bar{p}_0+\bar{E}\right)\hat{H}'_2\right]\over 
\frac{\alpha \bar{P}}{M}}
\left(n-n_0\right)\right)\right. \\
&\times&\left.
\hat{U}_G \rho_\phi(n_0)\rho_{q^0}(n_0) \hat{U}_G^\dagger
\exp\left(-i{\left[\left(\hat{p}_0+\bar{E}\right)^2+
2 \left(\bar{p}_0+\bar{E}\right)\hat{H}'_2\right]\over 
\frac{\alpha \bar{P}}{M}}
\left(n-n_0\right)\right) \right]{\rm Den}^{-1}.\nonumber
\end{eqnarray}

The value of the denominator actually does not change due to the boost, 
although its form changes. We will address this point later on.

To understand the covariance it is convenient to commute $\hat{U}_G$
with the exponential; let us therefore analyze the
product,
\begin{equation}
  B= \exp\left(i{\left[\left(\hat{p}_0+\bar{E}\right)^2+
2 \left(\bar{p}_0+\bar{E}\right)\hat{H}'_2\right]\over 
\frac{\alpha \bar{P}}{M}}
t_{\rm max}(n)\right)
\exp\left(i \frac{m \beta^2}{2} \hat{q}^0\right)
\exp\left(i\beta\hat{\pi}\hat{q}^0\right)\exp\left(-im\beta\hat{\phi}\right),
\end{equation}
where we have used the fact that,
\begin{equation}
\hat{U}_G=\exp\left(i \frac{m \beta^2}{2} \hat{q}^0\right)
\exp\left(i\beta\hat{\pi}\hat{q}^0\right)\exp\left(-im\beta\hat{\phi}\right),
\label{65}
\end{equation}
which can be shown using the Baker--Campbell--Hausdorff formula.

We wish to commute $\hat{U}_G$ to the left. We start by noting that in the
subspace of $\phi,\pi$, the variable $q^0$ behaves as an external
parameter, as if it were a classical time $t$. Following the
calculations of \cite{brho} for an ordinary quantum system one has,
\begin{equation}
  \exp\left(i\hat{H}'(t-t_0)\right) \hat{U}_{t_0}\psi(t_0) = 
\hat{U}_t \exp(i \hat{H} (t-t_0)\psi(t_0),
\end{equation}
which just states that the evolution of the Galilean transformed state
should coincide with the Galilean transform of the original evolved
state. That is,
\begin{equation}
\exp\left(i \hat{H}'(t-t_0)\right)\hat{U}_{t_0} \psi(t_0)=
\exp\left(i m {\beta^2 \over 2} (t-t_0)\right)\exp\left(i \beta
\hat{\pi}(t-t_0)\right) \hat{U}_{t_0} \exp\left(i \hat{H} 
(t-t_0)\right)\psi(t_0).
\end{equation}

We can therefore write for our system,
\begin{eqnarray}
P'\left(\phi \in \Delta \phi|q^0\in \Delta t\right)&=&\sum_n 
{\rm Tr}\left[\hat{P}_{\phi,q^0} 
\exp\left(i \left[ \frac{m \beta^2}{2} t_{\rm max}(n) +i\beta \hat{\pi}
    t_{\rm max}(n)\right]\right)
\exp\left(i {\left(\hat{p}_0+\hat{E}\right)^2\over \frac{\alpha
      \bar{p}}{M}}\left(n-n_0\right)\right) \hat{U}_G\right.
\nonumber\\
&\times& 
\exp\left(i \hat{H}_2 t_{\rm max}(n)\right) \rho_\phi(n_0) \rho_{q^0}(n_0)
\exp\left(-i \hat{H}_2 t_{\rm max}(n)\right)\hat{U}_G^\dagger
\exp\left(-i \frac{\left(\hat{p}_0+\hat{E}\right)^2}{\frac{\alpha
      \bar{p}}{M}}\left(n-n_0\right)\right)\nonumber\\
&\times& \left.
\exp\left(-i 
\left[ \frac{m \beta^2}{2} t_{\rm max}(n) +i\beta \hat{\pi}
    t_{\rm max}(n)\right]\right)
\right]{\rm
Den}^{-1}.\label{68}
\end{eqnarray}

We still need to commute $\exp\left(i
{\left(\hat{p}_0+\hat{E}\right)^2\over \frac{\alpha
\bar{p}}{M}}\left(n-n_0\right)\right)$ with $\hat{U}_G$. Noting that
the expression for $\hat{U}_G$ (\ref{65}) can be written as,
\begin{equation}
  U_G = \exp\left(i \left[\frac{\beta^2 m}{2} +\beta \hat{\pi}\right] \hat{q}_0 \right)\exp \left(-im\beta \hat{\phi}\right)
\end{equation}
we see that only the first term in the exponential has a non-trivial
commutator.

To proceed we note that
if one has two operators $\hat{A},\hat{B}$ such that
$[[\hat{A},\hat{B}],\hat{A}]=0$, one has that,
\begin{equation}
  e^{\hat{A}} e^{\hat{B}} = e^{\frac{1}{2} [\hat{A},\hat{B}]} 
e^{\hat{A}+\hat{B}}.
\end{equation}

If we now take $A = a(\hat{p}_0+\bar{E})^2$ and $B=b \hat{q}^0$ we
have the following identities,
\begin{equation}
\exp\left(ia \left(\hat{p}_0+\bar{E}\right)^2\right) \exp\left(i b\hat{q}_0\right) =
\exp\left(i a b \left(\hat{p}_0+\bar{E}\right)\right) \exp\left(ia
  \left(\hat{p}_0+\bar{E}\right)^2 +i b \hat{q}_0\right),
\end{equation}
and,
\begin{equation}
\exp\left(ib \hat{q}_0\right) 
\exp\left(ia\left(\hat{p}_0+\bar{E}\right)^2\right) = 
\exp\left(-iab\left(\hat{p}_0+\bar{E}\right)\right) 
\exp\left(-\frac{i}{6} ab^2\right) 
\exp\left(ia \left(\bar{p}_0+\bar{E}\right)^2+ib \hat{q}_0\right),
\end{equation}
therefore,
\begin{equation}
\exp\left(ia\left(\bar{p}_0+\bar{E}\right)^2\right)
\exp\left(ib\hat{q}_0\right) = 
\exp\left(2iab\left(\hat{p}_0+\bar{E}\right)\right) 
\exp\left(\frac{i}{6} ab^2\right) 
\exp\left(ib\hat{q}_0\right) 
\exp\left(ia\left(\hat{p}_0+\bar{E}\right)^2\right).  
\end{equation}

Taking $a={i (n-n_0) \over \frac{\alpha \bar{p}}{M}}$ and $b=
\frac{m}{2} \beta^2 +\beta\hat{\pi}$ and substituting
$\hat{p}_0+\bar{E}$ by $\bar{p}_0+\bar{E}$ in (\ref{68}) we 
finally have,
\begin{eqnarray}
P'\left(\phi \in \Delta \phi|q^0\in \Delta t\right)&=&\sum_n 
{\rm Tr}\left[\hat{P}_{\phi,q^0} \hat{U}_G
\exp\left( i { \left[ \left(\hat{p}_0+ \bar{E}\right)^2 +2
      \left(\bar{p}_0 +\bar{E}\right) \hat{H}_2\right]\over
    \frac{\alpha \bar{p}}{M}}(n-n_0)\right) 
\rho_\phi \rho_{q_0} \right.\\
&\times&\left. 
\exp\left(- i { \left[ \left(\hat{p}_0+ \bar{E}\right)^2 +2
      \left(\bar{p}_0 +\bar{E}\right) \hat{H}_2\right]\over
    \frac{\alpha \bar{p}}{M}}(n-n_0)\right)
\hat{U}_G^\dagger\right]
{\rm Den}^{-1}.
\end{eqnarray}

It is remarkable that all the terms involving $t_{\rm max}$ in (\ref{68})
have cancelled with the terms stemming from the commutation we just did.
We now can address the point we postponed before, namely the change in 
the denominator of the expression. Basically, a similar calculation to the
one we just did starting from (\ref{45}) and performing the commutations
shows that the denominator is actually invariant, using the cyclicity of the
trace and the fact that unlike the numerator, it does not involve the 
projector on the $\phi$ space.

Now since
\begin{equation}
  \hat{U}^\dagger_G \hat{P}_{\phi,q^0} \hat{U}_G = \hat{P}_{\phi -\beta q^0,q^0},
\end{equation}
we therefore have,
\begin{equation}
P'\left(\phi \in \Delta \phi|q^0 \in \Delta t\right)
=P\left(\phi -\beta q^0 \in \Delta \phi|q^0 \in \Delta t\right)
\end{equation}

Which shows that the conditional probability is Galilean invariant.

\section{Conclusions}

We have shown in a simple model that considering a quantum clock in
quantum mechanics leads to a modification of Schr\"odinger equation,
and that the resulting probabilities are Galilean invariant.  Since
the probabilities are invariant, then physical predictions from this
framework will also be invariant. In particular the rate of
decoherence predicted in \cite{deco1,njp,bh} should be invariant.
This is an interesting point since the rate of decoherence predicted
is proportional to the difference of energies of states of the system
in an basis of energy eigenstates. One could ask the question, how can this
formula be Galilean invariant since the energy is not?  The answer has
to do with the fact that in order to have an energy  basis as
the one assumed in the calculation (with discrete spectrum) one has to
consider systems analogous to a particle in a potential. In such
systems, at least if they are isolated, the difference between energy
levels is a Galilean invariant and therefore the decoherence rate is a
Galilean invariant.


It remains to be studied how the decoherence presented would transform
under Lorentz transformations. Milburn \cite{Milburn} recently studied
decoherence in a Lorentz invariant setting and his treatment could
provide a framework to analyze our proposal in some detail. This is
more problematic, since we are considering corrections to quantum
mechanics, and if one goes to the relativistic domain one first has to
contend with the usual difficulties of defining a relativistic quantum
mechanics. Although it appears that the use of a relational time could
yield a well defined theory, the details remain to be studied.

\section{Acknowledgments}
This work was inspired by questions by Ted Jacobson and Daniel
Sudarsky at the Ladek Zdroj meeting on theoretical physics, JP thanks
the organizers, Jerzy Kowalski-Glikman and Giovanni Amelino-Camelia
for hospitality.  This work was supported by grant NSF-PHY0244335 and
funds from the Horace Hearne Jr. Laboratory for Theoretical Physics
and the Abdus Salam International Center for Theoretical Physics.

\end{document}